\documentclass{elsart}

\usepackage{natbib}

\usepackage{amssymb}

\def\ltw{\>\hbox{\lower.25em\hbox{$\buildrel <\over\sim$}}\>}
\def\gtw{\>\hbox{\lower.25em\hbox{$\buildrel >\over\sim$}}\>}
\begin{document}

\begin{frontmatter}



\title{Particle Acceleration in the M87 Jet}


\author[jae]{Jean Eilek}, \author[peh]{Philip Hardee}, 
\author[al]{Andrei Lobanov}

\address[jae]{New Mexico Tech, Socorro, NM, USA}

\address[peh]{University of Alabama, Tuscaloosa, AL, USA}

\address[al]{Max-Planck-Institut f\" ur Radioastronomie, Bonn, Germany}

\begin{abstract}
The wealth of high quality data now available on the M87 jet inspired
us to carry out a detailed analysis of the plasma physical conditions
in the jet.  In a companion paper (Lobanov, Hardee \& Eilek, this
proceedings) we identify a double-helix structure within the jet, and
apply Kelvin-Helmholtz stability analysis to 
determine the physical state of the jet plasma. 
In this paper we treat the jet as a test case for {\it in
situ} particle acceleration.  We find that plasma turbulence is likely
to exist at levels which can maintain the energy of electrons radiating
in the radio to optical range, consistent with the broadband spectrum
of the jet.

\end{abstract}




\end{frontmatter}

\section{Introduction}

We know a great deal about the jet in M87.  Radio \citep{OHC} and
optical \citep{HST} images reveal ordered, filamentary structures whose 
synchrotron emission extends from radio at least to optical frequencies, and 
probably up to X-rays \citep{Xray}.  Multi-epoch studies detect 
relativistic proper motion, at $\gamma$ of a few\citep{Biretta}.

The jet begins by expanding uniformly, out to $\sim 2$ kpc (we assume an 
angle $\sim 40^{\circ}$ to the line of sight, and a distance of 17 Mpc).
   At that point, the location
of the bright knot A, it recollimates, and continues for another 1-2 kpc
before disrupting strongly and depositing its matter and energy in
the larger radio halo. 
The minimum pressure required to produce the synchrotron emission 
remains approximately constant during the expansion.
Because $p_{min}$ measures the energy density in relativistic electrons
and magnetic field, $p_{min} \propto u_e^{4/7} u_B^{3/7}$, we know that
these quantities do not decay adiabatically during the expansion.  This is
clear evidence that {\it in situ} energization is occuring.

The broadband, slice-integrated synchrotron spectrum is nearly 
constant along the jet. This suggests that the 
electron energy distribution  is also nearly constant along the jet,
and  supports the idea that it is
 maintained against losses by {\it in situ} energization.
We must recall, however, that the jet is not internally homogeneous.
The radio-bright filaments probably show us high-field regions.  
In addition, resolved two-dimensional images show 
that the optical knots are more concentrated than the radio knots. 

We have enough information about this jet to justify detailed modelling
of the turbulence and its effect on the particles. 
To be specific, we consider particle acceleration
by Alfvenic turbulence. 
  In this paper we summarize our analysis, which will be published
elsewhere in more detailed form. 

\section{Physical picture of the jet}

From the radio and optical data we determine the overall physical state of
the jet. In a companion paper
\citep{LHE} we show that the jet has an underlying double helix structure,
which can be traced in radio and optical emissivity from the jet origin
out to and past knot A.  The structure of the helical
filaments is consistent with elliptical and helical normal
modes, which could be excited by the Kelvin-Helmholtz (KH) instability.  Some
of the bright ``knots'' in the jet are the points where the two filaments
rotate into the line of sight.  Other bright regions are superimposed 
on this pattern, such as the inner knots D and F, and 
the more ordered complex at and beyond knot A.  We think knot A is a
shock, connected to the recollimation of the jet,
 with emissivity enhanced by shear-driven
turbulence.  Past knot A the helical mode grows to large
amplitudes, creating the distortions which ultimately disrupt the flow.

We emphasize that the jet is not axisymmetric, despite the very well
defined apparent ``edges'' to the flow inside of knot A.  
From the structure of the excited modes \citep{Hardee}, we
know the plasma
 has localized high-pressure regions, close to the surface, which
rotate through a helical pattern going along the jet.   We propose that
these regions, and other randomly placed local bright spots, are
regions of enhanced, shear-driven MHD turbulence.  We further speculate
that this turbulence accelerates relativistic particles {\it in situ}, 
creating the radio and optical knots in the jet. 

We can combine standard minimum-pressure analysis and KH 
stability analysis to estimate physical parameters within the jet.
From the former \citep[e.g.]{OHC}, we scale to $B \sim 100 \mu $G, and 
$p \sim 10^{-9}$dyn/cm$^2$.  From stability analysis \citep{LHE}, 
we estimate the internal Mach number of the jet to be a few, and the
specific enthalpy of the jet plasma to be $ h = \Gamma p / (\Gamma -1)
\rho c^2 \sim 1$.  Thus, the jet cannot contain only highly
relativistic plasma, and must have total $ p > p_{min}$.  

The jet must
be in pressure balance with its surroundings in order to apply KH analysis.
This means that the inner few kpc of the Virgo core [which contain the
inner radio lobes \citep{Hines} and the complex X-ray-loud plasma 
\citep{WY}] must be at a higher pressure than earlier, low-resolution
X-ray work suggested.  This is consistent with the dynamic appearance
of the central region \citep{WY,OEK}.  It is also consistent
with the jet power, which we know is at least $3 \times 10^{44}$erg/s 
\citep{OEK}, and significant to the energy budget of the 
Virgo core.

\section{Jet deceleration and plasma turbulence}

We model the bright knots as localized sites of Alfvenic
turbulence.  This turbulence will 
accelerate relativistic particles.  It will also trap
 the newly energized  particles, so that they 
must diffuse away from the hot spots.
The observation that the radio knots are more extended than the
optical knots can be explained if the diffusion time, $\tau_{diff}$,
is comparable to the optical synchrotron loss time, $\tau_{sy,o}$.  

We begin by determining the turbulence level necessary to contain
the optically loud electrons. 
 Let $\delta B^2 / 8 \pi$ be the energy
density in MHD turbulence, and let the turbulence have a
characteristic (outer) scale
$\lambda_t$.   We follow \citep{GJ} and estimate the cross-field diffusion
coefficient $D_o \sim ( c \lambda_t / 3) \delta B^2 / B^2$.  
We scale both the hot spot size  and
the turbulent scale  to 10 pc.  With these scalings, we estimate  
$\tau_{diff} \sim \tau_{sy,o}$ if $\delta B^2 / B^2 \sim 0.1$.

Is this a reasonable turbulence level for the M87 jet?  If the
turbulence is driven by shear instabilities, the energy must come from 
the jet power, $P_j = \dot M \gamma_j c^2 \left( 1 + h \right)$ (where $\dot M$
is the mass flux).  As friction and velocity
shear decelerate the jet, energy is lost at a rate $ d P_j / dz$.
  We assume a fraction $\epsilon $ of this goes to drive the
turbulence;  the rest goes directly to heating the jet and ambient plasmas. 
The most important  damping mechanism in this situation
 is  jet expansion (which 
``adiabatically'' damps the turbulence).  Balancing this against the
driving, with jet deceleration $d \gamma/ d z
 \sim 1$/kpc, we find a very modest $\epsilon \sim .01$ 
can produce the required level of turbulence.
We further note that wave-wave interactions, which modulate the 
turbulent spectrum and generate dissipative modes, probably set in at
$\delta B^2 / B^2 \sim 0.1$ \citep[e.g.]{Spang}.  This also agrees with
our estimate above. 

\section{Impact on relativistic particles}

We next  ask whether
this turbulence level can accelerate the electrons. 
Alfven wave acceleration proceeds {\it via} the cyclotron
resonance, which requires  the wavelength and  particle energy
to match, as $ \gamma / \lambda \simeq e B / 2 \pi m c^2$.  
Very small scales are involved:  wavelengths $\lambda \sim 10^{11}$ to
 $10^{14}$ cm resonate with radio-loud ($\gamma \sim 10^4$)
to  X-ray loud ($\gamma \sim 10^7$) electrons.
The acceleration rate is determined by the
turbulent energy at  resonant scales. 
We expect  turbulent power to cascade
to smaller scales, but the resultant wavenumber
spectrum, $W(k)$, is hard to predict.
We follow tradition by assuming $W(k) \propto k^{-m}$, with
total energy $\delta B^2 / 8 \pi = \int W(k) dk$.
Stochastic acceleration is described by a Fokker-Planck
equation; quasi-linear theory can be used
to find the momentum diffusion coefficient $D_p$, and from that $
\tau_{acc} \sim p^2 / D_p \propto \gamma^{2-m}$
  \citep[e.g.]{JAE}.   

Can such turbulent acceleration offset radiative losses in the
M87 jet?   Do the accelerated particles produce a synchrotron spectrum
like that observed?  We have
only begun this calculation.  Our preliminary work suggests that
wave spectra with $m \simeq 3/2$,  which is typical of MHD turbulence in
other settings \citep{Kraich}, will offset synchrotron losses
for electrons radiating at and below the optical range.   This is pleasing, because
it seems consistent with the observed synchrotron
spectrum, which curves slowly between
the radio to the optical, then steepens further while continuing
to X-rays\citep{OHC,Xray}.  

Our simple estimates must be extended before they can usefully
be tested against the data.  
The particle energy distribution must be determined, and
convolved with the magnetic field structure in the jet, in order to
predict a synchrotron spectrum which can be compared to  
observations.  We are in the midst of those calculations, and will
report their results in a future paper.




\begin{thebibliography}{xxx}






\bibitem{Biretta}  Biretta, J. A., Zhou, F. \& Owen, F. N., 1995, ApJ,
447, 582

\bibitem{JAE} Eilek, J. A. \& Hughes, P. E., 1991, in P. E. Hughes,
ed., {\it Beams and Jets in Astrophysics} (Cambridge:  CUP), 428 

\bibitem{GJ} Giacalone, J. \& Jokipii, J. R., 1999, ApJ, 204, 214,
and references therein.


\bibitem{Hardee}Hardee, P. E., 2000, ApJ, 533, 176

\bibitem{Hines} Hines, D. C., Owen, F. N. \& Eilek, J. A., 1989, ApJ,
347, 713

\bibitem{Kraich} Kraichnan, R. H., 1965, Phys Fluids, 8, 1385


\bibitem{LHE}Lobanov, A., Hardee, P. E. \& Eilek, J. A., these proceedings.



\bibitem{Xray} Marshall, H. L., Miller, B. P., Davis, D. S. etal,
2002, ApJ, 564, 683


\bibitem{OEK} Owen, F. N., Eilek, J. A. \& Kassim, N. E., 2000, ApJ,
543, 611

\bibitem{OHC}Owen, F. N., Hardee, P. E. \& Cornwall, T. J., 1989, ApJ,
340, 698

\bibitem{HST} Perlman, E. S., Biretta, J. A., Sparks, W. B. Macchetto,
D. F. \& Leahy, J. P., 2001, ApJ, 551, 206


\bibitem{Spang} Spangler, S. R. \& Sheerin, J. P., 1983, ApJ, 272, 273



\bibitem{WY} Young, A. J., Wilson, A. S. \& Mundell, C. G., 2002, ApJ,
579, 560


\end{thebibliography}
\end{document}